\documentstyle[preprint,prl,aps]{revtex}
\begin{document}
\def\sn2{$\sin^22\theta$}
\def\dm2{$\Delta m^2$}
\def\ch2{$\chi^2$}
\def\ltap{\ \raisebox{-.4ex}{\rlap{$\sim$}} \raisebox{.4ex}{$<$}\ }
\def\gtap{\ \raisebox{-.4ex}{\rlap{$\sim$}} \raisebox{.4ex}{$>$}\ }
\newcommand{\EQ}{\begin{equation}}
\newcommand{\EN}{\end{equation}}
\draft
\begin{titlepage}
\newpage
\preprint{\vbox{\baselineskip 10pt{
\hbox{Ref. SISSA 13/99/EP}
\hbox{February 1999}
\hbox{hep-ph/9903303}}}}
\vskip -0.4cm
\title{ \bf Enhancing the Seasonal Variation Effect in the Case
of the Vacuum Oscillation Solution of the Solar Neutrino Problem}

\author{M. Maris$^{a}$ and S.T. Petcov $^{b,c)}$\footnote{Also at:
Institute of Nuclear Research and Nuclear Energy, Bulgarian Academy
of Sciences, BG--1784 Sofia, Bulgaria.}}
\address{a) Osservatorio Astronomico di Trieste, I-34113 Trieste, Italy}
\address{b) Scuola Internazionale Superiore di Studi Avanzati, I-34013
Trieste, Italy}
\address{c) Istituto Nazionale di Fizica Nucleare, Sezione di Trieste, I-34013 
Trieste, Italy}  
\maketitle
\begin{abstract}
\begin{minipage}{5in}
\baselineskip 16pt
We study in detail the 
threshold energy dependence of the 
seasonal variation effect in the
energy integrated  
solar neutrino signal of the  
Super-Kamiokande detector in the case of 
the $\nu_{e}\leftrightarrow \nu_{\mu,\tau}$
vacuum oscillation (VO) solution of the solar 
neutrino problem. 
We show, in particular, that for the values of $\Delta m^2$ 
and $\sin^22\theta$ from the VO solution region,  
the predicted time and threshold e$^{-}$ energy
(${\rm T_{e,Th}}$)
dependence of the event rate
factorize to a high degree of accuracy.
As a consequence, the VO generated seasonal 
variation asymmetry is given by the product
of an time-independent function of ${\rm T_{e,Th}}$ 
and the standard geometrical asymmetry. 
For any given $\Delta m^2$ 
and $\sin^22\theta$ 
from the VO solution region  
there exists at least one value of 
${\rm T_{e,Th}}$ from 
the interval ${\rm (5 - 11)~MeV}$, for which  
the seasonal variation effect in the solar neutrino 
sample of events, formed by recoil electrons 
with kinetic energy ${\rm T_{e} \geq T_{e,Th}}$, 
is either maximal or very close to the maximal; 
it can vary dramatically with ${\rm T_{e,Th}}$.
One can effectively search for the VO induced 
seasonal effect by forming a set of 
samples of solar neutrino events, 
corresponding to a sufficiently large 
number of different values of 
${\rm T_{e,Th}}$ from the indicated interval
and by measuring the seasonal variation 
in each of these samples. Predictions for the 
magnitude of the seasonal effect 
in such samples are given for a 
large set of representative 
values of $\Delta m^2$ 
and $\sin^22\theta$ from the VO solution region.

\end{minipage}
\end{abstract}
\end{titlepage}
\newpage

\hsize 16.5truecm
\vsize 24.5truecm
\def\dm{$\Delta m^2$\hskip 0.1cm}
\def\dmsqua{$\Delta m^2$\hskip 0.1cm}
\def\sn{$\sin^2 2\theta$\hskip 0.1cm }
\def\snf{$\sin^2 2\theta$}
\def\trna{$\nu_e \rightarrow \nu_a$}
\def\trnm{$\nu_e \rightarrow \nu_{\mu}$}
\def\trns{$\nu_e \leftrightarrow \nu_s$}
\def\trnat{$\nu_e \leftrightarrow \nu_a$}
\def\trnmt{$\nu_e \leftrightarrow \nu_{\mu}$}
\def\trne{$\nu_e \rightarrow \nu_e$}
\def\trnst{$\nu_e \leftrightarrow \nu_s$}
\def\nue{$\nu_e$\hskip 0.1cm}
\def\numu{$\nu_{\mu}$\hskip 0.1cm}
\def\nutau{$\nu_{\tau}$\hskip 0.1cm}

\font\eightrm=cmr8
\def\aprle{\buildrel < \over {_{\sim}}}
\def\aprge{\buildrel > \over {_{\sim}}}
\renewcommand{\thefootnote}{\arabic{footnote}}
\setcounter{footnote}{0}
\tightenlines

\vskip 0.4cm
\leftline{\bf 1. Introduction}
\vskip 0.3cm
\indent The hypothesis that the solar 
neutrinos undergo vacuum oscillations
into muon and/or tau neutrinos,
$\nu_{e}\leftrightarrow \nu_{\mu,\tau}$,
when they travel from the Sun to the Earth 
\cite{Pont67,Pont57,BPet87,KP1,KP2,KP3,KP4} continues to be 
a viable and very appealing explanation \cite{SKSmyDPF99,BKS98} 
of the observed deficit 
of solar neutrinos 
\footnote{\tightenlines \footnotesize 
The possibility of solar $\nu_{e}\leftrightarrow \nu_{s}$
oscillations, $\nu_s$ being a sterile neutrino, is 
disfavored by the existing solar neutrino data 
\cite{KP2,KP4,BKS98}.} \cite{Cl,Kam,GALLEX,SAGE,SKSuzukinu98,JNB}.
The mean event rate data from the solar neutrino
experiments Homestake, Kamiokande, SAGE, GALLEX and
Super-Kamiokande, can be described in terms of two-neutrino
$\nu_{e}\leftrightarrow \nu_{\mu (\tau)}$ oscillations
if the values of the two parameters, $\Delta m^2$ and 
$\sin^22\theta$, characterizing  the oscillations,
lie in the region \cite{SKSmyDPF99,BKS98}: 

$$5\times10^{-11} {\rm eV}^2\ltap \Delta m^2 \ltap 5\times
10^{-10}~ {\rm eV}^2,~~\eqno(1)$$
$$0.6 \ltap \sin^22\theta \leq 1.0.~~~\eqno(2)$$

\noindent Although this result is obtained utilizing the standard
solar model predictions of 
ref. \cite{BP98} for the fluxes of 
the $pp$, $pep$, $^{7}$Be, $^{8}$B and CNO neutrinos, it is rather stable
with respect to variation of the fluxes within their estimated
uncertainty ranges \cite{KP4,BKS98}. If, for instance, one treats the total
$^{8}$B neutrino flux as a free parameter 
in the mean event rate data analysis,
the solution regions located within the limits determined
by eqs. (1) and (2) change somewhat their position 
and magnitude only.
In this case two new ``low $^{8}$B neutrino flux'' solution regions
appear \cite{KP4} (at 95\% C.L.) at $\sin^22\theta \gtap 0.6$ for
$\Delta m^2 \sim (3 - 4)\times
10^{-11}~ {\rm eV}^2$ and 
$\Delta m^2 \sim (4 - 8)\times
10^{-12}~{\rm eV}^2$ (for a recent analysis see \cite{BKS98}). 
For values of $\Delta m^2 \sim (4 - 8)\times
10^{-12}~{\rm eV}^2$ the effects of the vacuum
$\nu_{e}\leftrightarrow \nu_{\mu (\tau)}$ oscillations
on the flux of $^{8}$B neutrinos with energies
$E_{\nu} \gtap 5$ MeV, which is presently studied in the
Super-Kamiokande experiment and will be studied in the near future
with the SNO detector, are hardly observable.
This low $^{8}$B flux vacuum oscillation (VO) solution
can be tested \cite{KP4} in the future solar neutrino experiments 
BOREXINO, HELLAZ, etc., and we will not discuss it further.

  When the current Super-Kamiokande results on the recoil-$e^{-}$
spectrum \cite{SKSmyDPF99} are added in the analysis 
of the solar neutrino data,
the regions of values of 
$\Delta m^2$ and $\sin^22\theta$ for which
the vacuum oscillations provide 
a good quality
fit of the data, are considerably reduced and the best fit
is achieved for $\Delta m^2 \cong (4.3 - 4.4)\times
10^{-10}~{\rm eV}^2$ and $\sin^22\theta \cong 0.9$.
Actually, the present Super-Kamiokande data on the spectrum
of the recoil electrons from the solar ($^{8}$B) 
neutrino induced reaction $\nu + e^{-} \rightarrow \nu + e^{-}$
favors the  $\nu_{e}\leftrightarrow \nu_{\mu (\tau)}$ 
vacuum oscillation solution of 
the solar neutrino problem over the MSW solutions \cite{SKSmyDPF99}.

   A strong evidence (if not a proof) that the solar 
neutrinos take part in vacuum 
oscillations on the way to the Earth would be 
the observation of a seasonal variation effect 
which differs from the standard geometrical one (see the second and the 
third articles quoted in \cite{Pont67} as well as 
\cite{KP1,KP2,KP3,KP4,FVO98,MSVO98,GlVO98,BeVO98,BarVO98}). 
For values of $\Delta m^2$ from the VO solution region (1)
and neutrino energies ${\rm E_{\nu}} \gtap 5$ MeV,
the oscillation length in vacuum,
${\rm L_v = 4\pi E_{\nu}/\Delta m^2}$, 
is of the order of, or exceeds but not 
by a very large factor, the seasonal change of the distance
between the Sun and the Earth, ${\rm \Delta R = 2\epsilon R_0}$,
where $\epsilon = 0.0167$ is the ellipticity of the Earth 
orbit around the Sun and ${\rm R_0} = 1.496\times 10^8$ km
is the mean Sun-Earth distance. This leads to a noticeable 
dependence of the neutrino oscillation probability
on the Sun-Earth distance, which in turn can create a 
specific seasonal difference in the solar neutrino signals
in detectors like Super-Kamiokande, SNO and ICARUS.
This difference should be particularly large in the 
signals due to the 0.862 MeV $^{7}$Be neutrinos \cite{KP3},
to be studied with the BOREXINO detector. 
No analogous effect is predicted 
in the case of the MSW solution of 
the solar neutrino problem \cite{SPJR89}.  

  The seasonal effect due to the vacuum oscillations  
of solar neutrinos can be amplified or reduced by the standard 
geometrical one caused by the 6.68\% decrease 
of the solar neutrino flux at the Earth surface 
when the Earth moves from perihelion (taking place in January) 
to aphelion (reached in July)
on its orbit around the Sun \cite{KP3}. In general, the VO 
induced seasonal effect
in the solar neutrino signals in the Super-Kamiokande, 
SNO and ICARUS detectors is predicted to be relatively small 
for values of $\Delta m^2$ from the interval (1).
For the envisaged and already attained rather 
low threshold e$^{-}$-kinetic energy of detection 
of the recoil electrons 
in the Super-Kamiokande experiment 
${\rm T_{e,Th}} \cong 5$ MeV
\cite{SKSmyDPF99}, 
the relative seasonal change of the event rate caused by 
the vacuum oscillations alone cannot exceed
(even under the most favorable experimental conditions) 
approximately 10\% \cite{KP3} (see further). 
This is below the sensitivity currently 
reached in the search for this effect in the
Super-Kamiokande experiment.
For the $\nu_{e}\leftrightarrow \nu_{\mu (\tau)}$
oscillations of interest
the effect in the Super-Kamiokande detector 
is reduced, in particular, by the 
weak neutral current interaction contribution of the 
$\nu_{\mu (\tau)}$ to the solar neutrino 
induced event rate. Correspondingly, the
VO induced seasonal variation 
of the solar neutrino signal should be,
in principle, somewhat larger 
(by a factor $\sim 1.2$) in the charged current 
event rates in the SNO and ICARUS experiments.
Nevertheless, the increase is not 
dramatic and it is worthwhile considering possible strategies
which might lead to the enhancement of the seasonal effect
in the samples of solar neutrino events 
collected by these or other 
real time detectors studying the $^{8}$B neutrino flux.

     It was noticed in ref. \cite{KP3} that 
the magnitude of the seasonal effect generated
by the vacuum solar $\nu_e$ oscillations in the  
Super-Kamiokande solar neutrino signal,
and more generally in the solar neutrino signals 
in detectors based on the reaction
$\nu + e^{-} \rightarrow \nu + e^{-}$,
is very sensitive to the change 
of the threshold 
e$^{-}$-kinetic energy from 
${\rm T_{e,Th}} \cong 5$ MeV to 
${\rm T_{e,Th}} \cong 7$ MeV. 
The VO induced seasonal effect in the 
signals in the SNO and ICARUS detectors 
should exhibit similar
strong dependence on the chosen minimal value of the 
energy of the detected final
state e$^{-}$. 
In the present article we explore this observation. We 
study in detail the threshold energy dependence of the 
seasonal variation effect in the sample of 
solar neutrino events in the  
Super-Kamiokande detector in the case of 
the vacuum oscillation solution of the solar 
neutrino problem. 
We show, in particular, that for the 
values of $\Delta m^2$ 
and $\sin^22\theta$ from the VO solution region (1) - (2),  
the predicted time and threshold  
e$^{-}$ kinetic energy (${\rm T_{e,Th}}$)
dependence of the energy integrated 
solar neutrino induced event rate factorize.
The VO generated seasonal variation asymmetry
is given by a product of a function which depends on
${\rm T_{e,Th}}$ but does not depend on time,
and of the standard geometrical asymmetry. 
We show also that for any given $\Delta m^2$ 
and $\sin^22\theta$ 
from the VO solution region  
there exists at least one value of the threshold
energy ${\rm T_{e,Th}}$ from 
the interval ${\rm (5 - 11)~MeV}$, for which  
the seasonal variation effect in the solar neutrino 
sample of events, formed by recoil electrons 
with energy ${\rm T_{e} \geq T_{e,Th}}$, 
is either maximal or very close to maximal. This suggests
a possible strategy to search effectively
for the VO induced seasonal effect:  
it consists in forming a set of 
samples of solar neutrino events, 
corresponding to a sufficiently large 
number of different values of 
${\rm T_{e,Th}}$ from the above interval,
say, to ${\rm T_{e,Th} = 5;~6;~7;...;~11~MeV}$,
and by measuring the seasonal variation 
in each of these samples. Predictions for the 
magnitude of the seasonal effect 
in each of the indicated samples
for a large set of representative 
values of $\Delta m^2$ 
and $\sin^22\theta$ from the VO solution 
region are also given.

\vskip 0.3cm
\leftline{\bf 2. The Seasonal Variation Effect Observables}
\vskip 0.3cm
\indent We shall consider the simplest case of 
solar  $\nu_{e}\leftrightarrow \nu_{\mu (\tau)}$
oscillations generated by two-neutrino mixing.
The probability that a solar electron neutrino with energy $E_{\nu}$ will
not change into $\nu_{\mu (\tau)}$ on its way to the
Earth when $\nu_e\leftrightarrow\nu_{\mu (\tau)}$ 
oscillations take place, has the well-known form:
\vskip -0.4truecm
$${\rm P(\nu_e\rightarrow\nu_e; R(t), E_{\nu}) 
\equiv P(R(t)) \equiv P(t) = 1 - {1\over 2}
\sin^22\theta~\bigl[ 1 - \cos 2\pi {R(t)\over
L_v}\bigr]},\eqno(3)$$
\noindent
where ${\rm L_v = 4\pi E_{\nu}}/\Delta m^2$ is 
the oscillation length in
vacuum, 
$${\rm R(t) = R_0~\bigl[ 1 - \epsilon\cos 2\pi {t\over
T}}\bigr] \equiv {\rm R_0~\hat{R}(t)}, \eqno(4)$$
\noindent is the Sun--Earth distance at time t of the year,
T = 365.24 days, ${\rm R_0} = 1.496\times 10^8$ km,
${\rm \hat{R}(t)}$ and $\epsilon =
0.0167$ being the mean Sun--Earth distance, the Sun-Earth distance 
at time $t$ in A.U.  and the ellipticity of the
Earth orbit around the Sun. 
 
   It is not difficult to check that     
in the neutrino energy interval of interest,
${\rm E_{\nu} \cong (5.0 - 14.4)~MeV}$, 
the probability ${\rm P(t)}$ has,
as a function of ${\rm E_{\nu}}$, 
i) one minimum 
for values of $\Delta m^2 \cong (0.5 - 0.8)\times 10^{-10}~{\rm eV^2}$, 
ii) one maximum and one minimum for 
$\Delta m^2 \cong (0.8 - 2.0)\times 10^{-10}~{\rm eV^2}$, etc.;
for $\Delta m^2 \cong (4.3 - 4.4)\times 10^{-10}~{\rm eV^2}$ it has
4 maxima and 3 minima. If, for instance, 
$\Delta m^2 = 0.75\times 10^{-10}~{\rm eV^2}$,
${\rm P(t)}$ decreases monotonically for
${\rm E_{\nu}}$ rising
in the interval 
${\rm E_{\nu} = (5.0 - 9.05)~MeV}$, and 
increases with ${\rm E_{\nu}}$
for ${\rm E_{\nu} = (9.05 - 14.4)~MeV}$;
for $\Delta m^2 = 2.0\times 10^{-10}~{\rm eV^2}$
it decreases as ${\rm E_{\nu}}$ increases
in the two intervals
${\rm E_{\nu} = (5.0 - 8.04)~MeV}$ and
${\rm E_{\nu} = (12.06 - 14.4)~MeV}$, and 
increases with ${\rm E_{\nu}}$ in the 
interval ${\rm E_{\nu} = (8.04 - 12.06)~MeV}$.
As it is easy to see from eq. (3), the derivatives of 
${\rm P(t)}$ with respect to ${\rm E_{\nu}}$
and ${\rm R(t)}$ have opposite signs \cite{MSVO98}. 
Correspondingly, ${\rm P(t)}$ will decrease from January
to June in the neutrino energy intervals, in which 
${\rm P(t)}$ increases with ${\rm E_{\nu}}$, and it will
increase from January to June in the energy
intervals where ${\rm P(t)}$ decreases with ${\rm E_{\nu}}$,
the seasonal effect having 
opposite signs in the two cases.
There is no seasonal change of 
${\rm P(t)}$ in the points of the extrema. 
Since the magnitude of the seasonal variation effect 
due to the vacuum oscillations is determined
by the seasonal change of the 
oscillation probability ${\rm P(t)}$,      
it should be clear from the above discussion that 
for the VO solution values of $\Delta m^2$ from (1),  
the seasonal effect in the solar neutrino induced
event rate integrated over the entire energy interval 
${\rm E_{\nu} = (5.0 - 14.4)~MeV}$ can be 
reduced considerably as a result of 
the mutual compensation between the opposite sign
seasonal effects in the ${\rm E_{\nu}}$ subintervals   
where ${\rm P(t)}$ increases and decreases with ${\rm E_{\nu}}$. 
As we shall see, for certain values of $\Delta m^2$
the compensation in the signal  of the Super-Kamiokande
detector is practically complete.  
We shall demonstrate in what follows that the indicated 
partial or complete compensation can be avoided in at least one 
of the samples
of events corresponding to different minimal values of
the recoil-e$^{-}$ kinetic energy, i.e., threshold energies 
${\rm T_{e,Th}}$, from the interval 
${\rm T_{e} \cong (5 - 11)~MeV}$. 
  
    The solar neutrino induced event 
rate in the Super-Kamiokande detector
at time $t$ of the year can be written for fixed  
threshold energy ${\rm T_{e,Th}}$, 
$\Delta m^2$ and $\sin^22\theta$ in the form: 
$${\rm R(t, T_{e,Th}};\Delta m^2,\theta) \equiv
{\rm R(t, T_{e,Th}}) =~ <~{\rm {1\over{\hat{R}^{2}(t)}}}~ 
\left [ {\rm r_{\nu} + (1 - r_{\nu})P(t)} \right ]~ >~,\eqno(5a)$$

\noindent where by the ``average'' of a quantity X, ${\rm < X >}$, we 
shall understand in what follows
$$< X >~ = {{\rm F(B)\over
{R_{0}^{2}}}}~
\int\limits_{{\rm T_{e,Th}}}^{ }
~{\rm dT_{e}}~
\int\limits_{{\rm T}_e(1 + {m_e\over 2{\rm T}_e})}^{ }
~{\rm dE_{\nu}}~
{\rm X~n(E_{\nu})}~{\rm (d\sigma (\nu_{e}e^{-})/dT_e)}.~\eqno(5b)$$

\noindent Here  ${\rm F(B)}/ {\rm R_{0}^{2} 
\equiv \bar{\Phi} (B)}$ is the mean annual 
$^{8}$B neutrino flux at the Earth surface,
${\rm n(E_{\nu})}$ is the normalized spectrum of $^{8}$B neutrinos,
$\int\limits_{0}^{14.4~{\rm MeV}}{\rm n(E_{\nu})dE_{\nu}} = 1$,
${\rm d\sigma (\nu_{l}e^{-})/dT_e}$ is the differential 
cross--section of the process $\nu_{l} + e^{-}
\rightarrow \nu_{l} + e^{-}$, $l=e,\mu~(\tau)$, and
${\rm r_{\nu} = 
(d\sigma (\nu_{\mu}e^{-})/dT_e)/ (d\sigma (\nu_{e}e^{-})/dT_e)}$.
In the neutrino energy interval of interest one has 
${\rm r_{\nu}} \cong (0.155 - 0.160)$. 
Expression (5b) is valid for ideal 
e$^{-}$ detection efficiency and energy resolution of the
Super-Kamiokande detector
\footnote{\tightenlines \footnotesize 
Obviously, expressions (5a) and (5b) 
and the results based on them we shall obtain, 
are valid for any  other ``ideal'' 
detector utilizing the $\nu + e^{-}
\rightarrow \nu + e^{-}$ reaction for 
detection of the $^{8}$B neutrinos. In this sense 
our results can serve as a  
guidance for the expected ${\rm T_{e,Th}}-$dependence
of the magnitude of the seasonal effect
in any detector of this type.}.
It can be trivially modified to include
the latter.  
The time-dependent quantities
in the expression
for ${\rm R(t, T_{e,Th})}$ are the Sun-Earth 
distance, i.e., the geometrical factor 
(expressed in A.U.), 
${\rm \hat{R}^{-2}(t)}$, 
and the probability ${\rm P(t)}$.

   In what follows we shall consider the event rate
${\rm R(t; T_{e,Th})}$, averaged over a time interval
with a central point ${\rm t = t_c}$ 
and width ${\rm \Delta t}$,
${\rm R(t_c, \Delta t; T_{e,Th})}$. 
The point ${\rm t_c}$ can be the time at 
which the Earth reaches the perihelion, ${\rm t_c = t_p = 0}$, or aphelion, 
${\rm t_c = t_a = T/2}$, or any other chosen time of the year. 
It is convenient to choose ${\rm t_p = 0 \leq t_c \leq t_a = T/2}$ 
since ${\rm R(t_c) = R(T - t_c)}$.
The seasonal effect due to VO should exhibit
a similar symmetry.
We shall present results for 
${\rm t_c = t_p;~t_p + T/12;~t_p + 2T/12;...;~t_a}$ and a width of one month, 
${\rm \Delta t = T/12}$. In their analysis of the solar neutrino data
the Super-Kamiokande collaboration is utilizing time
bins having a width of 1.5 months, ${\rm \Delta t = T/8}$ \cite{SKSmyDPF99}. 
As we shall see, the choice of ${\rm \Delta t \gtap T/4}$ tends to 
suppress the seasonal variation effect.

   Obviously, the time-averaged event rate
${\rm R(t_c, \Delta t; T_{e,Th})}$ is determined by the
time-averaged probability ${\rm P(t)}$. 
It proves useful to consider two different types of time-averaging. 
The first includes the effect of the geometrical factor,
$${\rm \bar{P}_{GF}(t_c,\Delta t) = {1\over {\Delta t}}}~
  \int\limits_{{\rm t_c - {1\over {2}} \Delta t}}
^{{\rm t_c + {1\over {2}} \Delta t}}~{\rm dt~
{P(t)\over{\hat{R}^{2}(t)}}}~,~\eqno(6)$$

\noindent while in the second the geometrical factor effect
is essentially eliminated:  
$${\rm \bar{P}_{NGF}(t_c,\Delta t)} = 
{{\rm \bar{P}_{GF}(t_c,\Delta t)\over {\kappa (t_c,\Delta t)}}}~,
~~\eqno(7)$$

\noindent where
$${\rm \kappa (t_c,\Delta t) = 
{1\over {\Delta t}}}~
  \int\limits_{{\rm t_c - {1\over {2}} \Delta t}}
^{{\rm t_c + {1\over {2}} \Delta t}~dt}~
{{\rm dt\over{\hat{R}^{2}(t)}}}~ 
= {\rm 1 +  2\epsilon \cos \left (2\pi {t_c\over T} \right )~
{{\sin 2\pi \Delta t/(2T)}\over
{2\pi \Delta t/(2T)}}} + O(\epsilon^2),~\eqno(8)$$
 
\noindent and we have used eq. (4) and the fact that
terms $\sim \epsilon ^2 \cong 2.8\times 10^{-4}$ are beyond
the sensitivity of the operating and planned solar neutrino
experiments and can be neglected.
It follows from eqs. (4) and (6) - (8) that the one year 
averaged probabilities ${\rm \bar{P}_{GF}(t_c,T)}$
and ${\rm \bar{P}_{NGF}(t_c,T)}$ practically coincide:
$${\rm \bar{P}_{GF}(t_c,T)} = 
{\rm \bar{P}_{NGF}(t_c,T)} + O(\epsilon^2).~\eqno(9)$$

\noindent The same result holds for the average probabilities
independently of ${\rm \Delta t}$ for
${\rm t_c = T/4;~3T/4}$, i.e., at the spring and autumn 
equinoxes:
${\rm \bar{P}_{GF}(t_c = T/4~(3T/4), \Delta t}) = 
{\rm \bar{P}_{NGF}(t_c = T/4~(3T/4), \Delta t}) + O(\epsilon^2)$.

  The expression for the average event rate in the time interval
${\rm [t_c - \Delta t/2,t_c + \Delta t/2]}$, in which the effect of the
geometrical factor is essentially excluded,
${\rm R_{NGF}(t_c, \Delta t; T_{e,Th})}$, can   
formally be obtained from eqs. (5a) - (5b) 
by replacing the probability ${\rm P(t)}$
with the probability ${\rm \bar{P}_{NGF}(t_c,\Delta t)}$ and
by setting the factor ${\rm \hat{R}^{-2}(t)}$ to 1:
$${\rm R_{NGF}(t_c, \Delta t; T_{e,Th})} = 
~ <~{\rm r_{\nu} + 
(1 - r_{\nu})}{\rm \bar{P}_{NGF}(t_c,\Delta t)}~>~.\eqno(10)$$

\noindent It is easy to see from eqs. (5) - (8) and (10)
and the above remark that
for the average event rate containing
the contribution of the geometrical factor,
${\rm R_{GF}(t_c, \Delta t; T_{e,Th})}$, one has:
$${\rm R_{GF}(t_c, \Delta t; T_{e,Th})} = 
{\rm \kappa (t_c,\Delta t)}~
 {\rm R_{NGF}(t_c, \Delta t; T_{e,Th})}.~\eqno(11)$$

\noindent The one year averaged event rates 
${\rm R_{GF}(t_c, T; T_{e,Th})}$ and
${\rm R_{NGF}(t_c, T; T_{e,Th})}$ coincide up to corrections
$\sim \epsilon^2$. For ${\rm t_c}$ corresponding to the perihelion, 
${\rm t_c = 0}$, we shall denote them both by
\footnote{\tightenlines \footnotesize Note, that the perihelion was reached 
on  2, 5, and 4 of January 
in 1997, 1998 and 1999,
respectively.}
${\rm R(T_{e,Th})}$:
${\rm R(T_{e,Th})} \equiv   
{\rm R_{NGF}(0, T; T_{e,Th})} \cong {\rm R_{GF}(0, T; T_{e,Th})}$.

  We shall analyze in what follows two observables: the ratio
$${\rm N_{i}(t_c, \Delta t; T_{e,Th}) = 
  {R_{i}(t_c, \Delta t; T_{e,Th})\over{\rm R(T_{e,Th})}}}~,
~{\rm i=NGF,~GF},~\eqno(12)$$

\noindent and the related seasonal variation asymmetry,
$${\rm A^{seas}_{i}(t_c, \Delta t; T_{e,Th})} = 
{\rm N_{i}(t_c, \Delta t; T_{e,Th}) -  
 N_{i}(t_c + T/2, \Delta t; T_{e,Th})~,~i= NGF,~GF}.~\eqno(13)$$

\noindent Since ${\rm N_{i}(t_c + T/2, \Delta t; T_{e,Th}) =
N_{i}(T/2 - t_c, \Delta t; T_{e,Th})}$, 
it is sufficient to consider 0${\rm \leq t_c \leq T/4}$
in the case of the asymmetry.
Note that both observables (12) and (13)
do not depend on the total flux of $^{8}$B neutrinos. 
The observable ${\rm N_{NGF}(t_c, \Delta t; T_{e,Th})}$
coincides with
the one introduced in ref. \cite{KP3}.
The seasonal variation asymmetries which have been
analyzed in refs. \cite{FVO98,MSVO98,GlVO98,BeVO98,BarVO98}
are analogous to, but differ somewhat from,
${\rm A^{seas}_{NGF(GF)}(t_c, \Delta t; T_{e,Th})}$.

   In the absence of neutrino oscillations ${\rm P(t)} = 1$ and we have:
${\rm N_{NGF}(t_c, \Delta t; T_{e,Th})} = 1$,
${\rm N_{GF}(t_c, \Delta t; T_{e,Th}) = \kappa (t_c,\Delta t)}$,
${\rm A^{seas}_{NGF}(t_c, \Delta t; T_{e,Th})} = 0$, and 
${\rm A^{seas}_{GF}(t_c, \Delta t; T_{e,Th})} =
{\rm A^{seas}_{geom}(t_c, \Delta t)}$, where 
${\rm A^{seas}_{geom}(t_c, \Delta t)}$ is the seasonal asymmetry 
of purely geometrical origin,
$${\rm A^{seas}_{geom}(t_c, \Delta t) = 
\kappa (t_c,\Delta t) - \kappa (t_c + T/2,\Delta t)}  
= 4\epsilon \cos \left ( 2\pi {\rm {t_c\over T}}\right )~
{\rm {{\sin2\pi \Delta t/(2T)}\over
{2\pi \Delta t/(2T)}} + O(\epsilon ^2)}.~\eqno(14)$$

\noindent For ${\rm 0 \leq t_c < T/4}$ (and ${\rm \Delta t < T}$)
the geometrical asymmetry is positive:
${\rm A^{seas}_{geom}(t_c, \Delta t)} > 0$.

  As can be shown, the following simple relations hold true up to
corrections $\sim 10^{-3}$ if the solar $\nu_e$ take part in two-neutrino
oscillations $\nu_{e}\leftrightarrow \nu_{\mu (\tau)}$ with 
$\Delta m^2 \ltap 5\times 10^{-10}~{\rm eV}^2$:
$${\rm N_{NGF}(t_c, \Delta t; T_{e,Th}) =
 1 + {1\over 2}~W(T_{e,Th})~A^{seas}_{geom}(t_c, \Delta t) 
+ O(10^{-3})},\eqno(15)$$
$${\rm N_{GF}(t_c, \Delta t; T_{e,Th}) 
= N_{NGF}(t_c, \Delta t; T_{e,Th}) + 
{1\over 2}A^{seas}_{geom}(t_c, \Delta t) + O(10^{-3})} $$
$${\rm = 1 + {1\over 2}~\left ( 1 + W(T_{e,Th})\right )~
A^{seas}_{geom}(t_c, \Delta t) 
+ O(10^{-3})},\eqno(16) $$
$${\rm A^{seas}_{NGF}(t_c, \Delta t; T_{e,Th}) =
 W(T_{e,Th})~A^{seas}_{geom}(t_c, \Delta t) 
+ O(10^{-3})},\eqno(17)$$
$${\rm A^{seas}_{GF}(t_c, \Delta t; T_{e,Th})} =
{\rm A^{seas}_{NGF}(t_c, \Delta t; T_{e,Th})} +
{\rm A^{seas}_{geom}(t_c, \Delta t) + O(10^{-3})}$$
$${\rm = \left ( 1 + W(T_{e,Th})\right )~A^{seas}_{geom}(t_c, \Delta t) 
+ O(10^{-3})}.\eqno(18)$$

\noindent The factor ${\rm W(T_{e,Th})}$ in eqs. (15) - (18) does not
depend on the time variables $t_c$ and  ${\rm \Delta t}$ and is given by:
$${\rm W(T_{e,Th}) = {1\over 2}\sin^22\theta~
{{<~(1 - r_{\nu})\pi {R_0\over L_v}~\sin 2\pi {R_0\over L_v}~>}\over
{<~r_{\nu} + (1 - r_{\nu}) P(R_0)~>}}}~,\eqno(19)$$
 
\noindent where ${\rm P(R_0)}$ is the one year average probability, 
i.e., ${\rm P(R(t))}$, eq. (3), in which the Sun-Earth distance 
${\rm R(t)}$ is replaced 
with the mean Sun-Earth distance ${\rm R_0}$.
In the case of absence of oscillations we have 
$\sin^22\theta = 0$ and/or $\Delta m^2 = 0$, and therefore 
${\rm W(T_{e,Th})} = 0$.
 
  The relations (15) - (18) can be derived using the following 
observations. For $\Delta m^2 \ltap 5\times 10^{-10} {\rm eV}^2$
and ${\rm E_{\nu}} \geq 5~$MeV we have 
$2\pi \epsilon {\rm R_0/L_v} \ltap 0.63$. Correspondingly,
one can expand the probability ${\rm P(R(t))}$ in power series of
${\rm x(t) = 2\pi \epsilon R_0/L_v\cos 2\pi t/T}$ \cite{KP3}. 
In this way one obtains:
$${\rm P(R(t))} = {\rm P(R_0) + \Delta P(t)},~~\eqno(20)$$

\noindent where 
 $${\rm \Delta P(t) = {1\over 2}\sin^22\theta 
\left [ (x(t) - {1\over 6}x^{3}(t) +...)\sin 2\pi {R_0\over L_v}
- {1\over 2}(x^{2}(t) - {1\over 12}x^4(t) + ...)\cos 2\pi {R_0\over L_v}
\right ]}.\eqno(21)$$

\noindent For $\Delta m^2 \leq 10^{-10}~{\rm eV}^2$ and 
${\rm E_{\nu}} \geq 5~$MeV one finds   
${\rm x(t)} \leq 0.13$. Thus, the term linear in 
${\rm x(t)}$ gives the dominant contribution in
${\rm \Delta P(t)}$: for the contribution of the 
quadratic term, for instance, we get
${\rm 0.25 x^{2}(t) \leq 4.3\times 10^{-3}}$.  
Keeping only the term linear in 
${\rm x(t)}$ in the expression for
${\rm \Delta P(t)}$ one arrives at the relations
(15) - (19). The linear approximation, as 
numerical studies we have performed
showed, turns out to be equally accurate for   
$10^{-10} {\rm eV}^2 < \Delta m^2 
\leq 5\times 10^{-10}~{\rm eV}^2$.
The reason for this somewhat unexpected
result lies in the fact that 
for $\Delta m^2 \geq 10^{-10}~{\rm eV}^2$ and 
${\rm E_{\nu}} \sim (5 - 14)~$MeV,
the argument of the $t-$independent sine 
and cosine functions in ${\rm \Delta P(t)}$ is 
relatively large: 
$2\pi {\rm R_0/L_v} \geq (2.7 - 7.5)$.
Thus, for $\Delta m^2 \geq 10^{-10}~{\rm eV}^2$,
the functions ${\rm (2\pi \epsilon R_0/ L_v)^{n} \sin 2\pi R_0/ L_v}$
and ${\rm (2\pi \epsilon R_0/ L_v)^{n} \cos 2\pi R_0/ L_v}$ with
$n \geq 2$ are fastly oscillating
functions of ${\rm E_{\nu}}$ and the integration
over the neutrino energy renders them negligible:
$$<{\rm {1\over {2(n!)}} (2\pi \epsilon R_0/ L_v)^{n} \sin (2\pi R_0/ L_v)> 
~\sim O(10^{-3}),~~n} \geq 3,~\eqno(22a)$$ 
 $$<{\rm {1 \over {2(n!)}} (2\pi \epsilon R_0/ L_v)^{n} 
\cos (2\pi R_0/ L_v)>~ \sim O(10^{-3}),~~n} \geq 2.~\eqno(22b)$$

   Relations (17) and (18) imply that to a 
high degree of accuracy the VO generated 
seasonal variation asymmetries 
${\rm A^{seas}_{NGF,GF}(t_c, \Delta t; T_{e,Th})}$
are proportional to the geometrical one, 
${\rm A^{seas}_{NGF(GF)}(t_c, \Delta t; T_{e,Th})} \sim
{\rm A^{seas}_{geom}(t_c, \Delta t)}$. 
Actually, the time ($t_c$ and ${\rm \Delta t}$) and the energy
dependencies of the observables 
${\rm N_{NGF,GF}(t_c, \Delta t; T_{e,Th})}$ and
${\rm A^{seas}_{NGF,GF}(t_c, \Delta t; T_{e,Th})}$
factorize: the whole time dependence is
contained in ${\rm A^{seas}_{geom}(t_c, \Delta t)}$ 
and is given
with a high precision by the factor 
$${\rm f(t_c, \Delta t)} = \cos \left ( 2\pi {\rm {t_c\over T}}\right )~
{\rm {{\sin2\pi \Delta t/(2T)}\over
{2\pi \Delta t/(2T)}}},~\eqno(23)$$

\noindent while the function ${\rm W(T_{e,Th})}$
carries all the information 
about the energy dependence.
Depending on the values of $\Delta m^2$ from the interval
(1) and on ${\rm T_{e,Th}}$, 
the function ${\rm W(T_{e,Th})}$ can be positive, negative or
zero for ${\rm E_{\nu} \cong (5.0 - 14.4)~MeV}$; 
and $|{\rm W(T_{e,Th})}|$ can be greater or 
smaller than 1 (see Table III).
Correspondingly, the VO generated and the geometrical seasonal
asymmetries ${\rm A^{seas}_{NGF}}$ and ${\rm A^{seas}_{geom}}$ 
can have the same or opposite signs and the former can be larger or smaller
in absolute value than the latter \cite{KP3}. 
The asymmetry which contains both the VO induced seasonal effect 
and the effect due to the geometrical factor,
${\rm A^{seas}_{GF}}$, is just equal to the sum 
of the seasonal asymmetry due to the VO only and of
the geometrical asymmetry, eq. (18). 
The asymmetry ${\rm A^{seas}_{GF}}$
can be close to zero due to mutual cancelation between
${\rm A^{seas}_{NGF}}$ and ${\rm A^{seas}_{geom}}$ \cite{KP3}.
This implies, in particular, that one cannot have
simultaneously 
$|{\rm A^{seas}_{NGF}| \ll A^{seas}_{geom}}$ and 
$|{\rm A^{seas}_{GF}| \ll A^{seas}_{geom}}$
for the same set of values of the parameters. However,
one of the two asymmetries, 
${\rm A^{seas}_{NGF}}$ or ${\rm A^{seas}_{GF}}$, 
can be strongly suppressed, while the other can have 
observable values. All these 
possibilities are realized for the values of
$\Delta m^2$, $\sin^22\theta$, ${\rm E_{\nu}}$ and ${\rm T_{e,Th}}$ 
of interest (see further and Figs. 1 and 2). 

    Let us note that in ref. \cite{FVO98} a Fourier 
analysis of the predicted 
time dependence of the event rate in, e.g., the 
Super-Kamiokande detector  
in the case of the VO solution of the solar
neutrino problem was performed. The coefficients
in the corresponding Fourier expansion
were found to be expressed in terms of Bessel
functions. The authors of ref. \cite{FVO98} noticed
that the constant and the first harmonic term dominate 
in the expansion. The origin of this 
result becomes clear from
eqs. (15) - (18) and (22a) - (22b). 

   We expect eqs. (15) - (18) and 
the above conclusions to be valid for the 
charged current solar neutrino event 
rates and the corresponding 
seasonal variation asymmetries 
to be measured with, e.g., the SNO and ICARUS detectors.
The expressions for the observables (12) and (13)
and for the function (19) for the SNO detector can be obtained 
simply by modifying the definition of 
the ``average'' in eqs. (5b) 
and of the averaged quantities 
in eqs. (10) and (19): one has to set 
${\rm r_{\nu} = 0}$  and replace in (5b)
${\rm d\sigma (\nu_{e}e^{-})/dT_e}$ with
${\rm d\sigma (\nu_{e} d \rightarrow e^{-} \rm pp)/dT_e}$ 
- the differential cross-section of the charged 
current reaction $\nu_{e} + {\rm d} \rightarrow e^{-} + {\rm p + p}$,
changing as well the lower limit of integration over ${\rm E_{\nu}}$ to 
${\rm T_{e}} + 1.44~$MeV. It should also be emphasized 
that our general results (15) - (19)
will not change if one includes in the analysis
the e$^{-}-$detection efficiency and 
energy resolution 
of a specific detector 
as long as they do not vary with time.  

  Let us note that in the case of averaging over a 
time period not exceeding two months,
${\rm \Delta t \leq T/6}$, we have
$$\kappa (t_c,\Delta t) 
= 1 +  2\epsilon \cos \left ( 2\pi {t_c\over T} \right )
 + O(\sim 1.7\times 10^{-3}),~\eqno(24)$$

$${\rm A^{seas}_{geom}(t_c, \Delta t) = 
4\epsilon \cos \left ( 2\pi {t_c\over T} \right )} + 
O(\sim 3.4\times 10^{-3}) 
,~\eqno(25)$$

\noindent and both quantities are practically
${\rm \Delta t}-$independent. The geometrical asymmetry 
is maximal if $t_c$ is chosen to be the time when
the Earth is at perihelion:
${\rm A^{seas}_{geom}(t_c = 0, \Delta t \leq T/6)} 
\cong 4\epsilon =$ 6.68\%. As it follows from eqs. (14) - (18),
averaging over a time interval exceeding 3 months,
${\rm \Delta t \gtap T/4}$, suppresses the seasonal variation
effect.

\vskip 0.3cm
\leftline{\bf 3. Enhancing the Seasonal Variation Effect}
\vskip 0.3cm
\indent  We have performed numerical calculations of the observables
(12) and (13) for a large representative 
set of values of the parameters
$\Delta m^2$ and $\sin^22\theta$ from the VO solution region
(1) - (2), which are given in Table I. Some of our results are 
presented in Tables I - III and in Figs. 1 - 3.
The threshold energy dependence of 
the event rate ratio (12), 
${\rm N_{NGF}(t_c, \Delta t; T_{e,Th})}$, is shown 
for six selected values of 
$\Delta m^2$ and $\sin^22\theta$ from the set 
and for ${\rm t_c = 0~(perihelion);~T/12;~ 2T/12;...;
~t_a = T/2~(aphelion)}$ and  
${\rm \Delta t = T/12}$ in Fig. 1.
The perihelion-aphelion asymmetries 
${\rm A^{seas}_{NGF,GF}(t_p = 0, \Delta t = T/12; T_{e,Th})
\equiv A^{seas}_{NGF,GF}(T_{e,Th})}$
(see eq. (13)) are plotted as functions of ${\rm T_{e,Th}}$
in Fig. 2 for all the values of $\Delta m^2$ and $\sin^22\theta$
given in the first column of Table I.
As Figs. 1 and 2 demonstrate, 
depending on  
$\Delta m^2$ and $\sin^22\theta$, 
for ${\rm T_{e,Th} \cong 5~MeV}$ 
the seasonal effect due to VO 
can be maximal or close to the maximal (Figs. 1a and 2a, 
${\rm \Delta m^2 = 0.5\times 10^{-10}~eV^2}$,
$\sin^22\theta = 0.9$),
or can be strongly suppressed (Figs. 1b, 1e, 1f and 2a, 2c,
${\rm \Delta m^2 = 0.7;~2.5;~4.4~\times 10^{-10}~eV^2}$).
With the increase of ${\rm T_{e,Th}}$ from the value of 
$\sim 5$ MeV, the VO induced
seasonal effect can change drastically.
For ${\rm \Delta m^2 = 0.9\times 10^{-10}~eV^2}$ 
and $\sin^22\theta = 1.0$ 
(Figs. 1c and 2b), for instance,
it is negative and close to maximal
in absolute value at ${\rm T_{e,Th} \cong 5~MeV}$, goes through
zero at ${\rm T_{e,Th} \cong 8.5~MeV}$ and is positive for larger
values of ${\rm T_{e,Th}}$ reaching a maximum
at ${\rm T_{e,Th} \cong 12~MeV}$. If, however,
${\rm \Delta m^2 = 0.4\times 10^{-10}~eV^2}$, 
it decreases monotonically with 
the increase of ${\rm T_{e,Th}}$ (Fig. 2a), while 
for ${\rm \Delta m^2 = 1.3\times 10^{-10}~eV^2}$ 
the effect is negative and hardly observable
at ${\rm T_{e,Th} \cong 5~MeV}$, but increases 
in absolute value reaching a maximum 
at ${\rm T_{e,Th} \cong 10.9~MeV}$,
${\rm A^{seas}_{NGF}(T_{e,Th} \cong 10.9~MeV)} \cong - 14.1\% $
(Figs. 1d and 2c).

   In the case of relatively large values of 
${\rm \Delta m^2 \cong (2.0 - 4.4)\times 10^{-10}~eV^2}$
the seasonal effect due to VO is relatively small or negligible for 
${\rm 5~MeV \leq T_{e,Th} \ltap 7.5~MeV}$ (Figs. 1e, 1f and 2c).
For ${\rm \Delta m^2 \cong 4.4\times 10^{-10}~eV^2}$ 
and $\sin^22\theta \cong 0.9$ the VO 
induced seasonal asymmetry increases with 
${\rm T_{e,Th}}$ reaching a maximum at 
${\rm T_{e,Th} \cong 10.1~MeV}$, 
${\rm A^{seas}_{NGF}(T_{e,Th} = 10.1~MeV)} \cong  14.8\% $,
then decreases to zero at
${\rm T_{e,Th} \cong 12.2~MeV}$ and becomes negative for
larger values of ${\rm T_{e,Th}}$.
The geometrical effect amplifies the asymmetry 
in the region of the maximum and 
${\rm A^{seas}_{GF}(T_{e,Th} = 10.1~MeV)} \cong  21.5\% $
(Fig. 2c). At ${\rm T_{e,Th} \cong 9.5~(11.0)~MeV}$, we have for
the indicated values of ${\rm \Delta m^2}$ and $\sin^22\theta$:
${\rm A^{seas}_{NGF}} \cong  13.0~(10.0)\% $.
Including the realistic experimental 
conditions in the calculations of the asymmetry 
will reduce its magnitude, but not considerably -
by $\sim (1 \div 2)\%$.
The change of ${\rm T_{e,Th}}$ from 5 MeV
to 10.1 MeV will increase the statistical error
of the measured value of ${\rm A^{seas}_{GF}(T_{e,Th})}$ 
(see Table I). Nevertheless, values of 
${\rm A^{seas}_{NGF(GF)}(T_{e,Th})} \sim 15~(21)\% $,  
as the one obtained above for 
${\rm \Delta m^2 \cong 4.4\times 10^{-10}~eV^2}$
at ${\rm T_{e,Th} \cong 10.1~MeV}$, 
can already be tested with the Super-Kamiokande
detector \cite{SKSmyDPF99}. 

   In Table II we give the numerical values of 
${\rm A^{seas}_{NGF}(T_{e,Th})}$ at
${\rm T_{e,Th} =}$ 5;~6;~7;~8;~9;~10;~11~MeV 
for the chosen representative set of values of $\Delta m^2$ and 
$\sin^22\theta$ from the VO solution region.
For each pair of the latter,
the values of ${\rm A^{seas}_{NGF}(T_{e,Th})}$ 
corresponding to a maximum of $|{\rm A^{seas}_{NGF}(T_{e,Th})}|$ 
in the interval ${\rm T_{e,Th} = (5 - 11)~MeV}$,
${\rm A^{seas,max}_{NGF}}$,
and the ${\rm T_{e,Th}}$ at which the 
maximum is reached, ${\rm T^{max}_{e,Th}}$, are also given.
For the set of values of neutrino parameters we have considered, 
the minimal ${\rm A^{seas,max}_{NGF}}$ is 2.2\%;
for most of the values, however, 
$|{\rm A^{seas,max}_{NGF}}| \gtap 4.5\%$, with the maximal 
values of the asymmetry 
reaching $\sim (10 \div 15)\%$. 
Our results show, in particular,
that for any given $\Delta m^2$ from the VO solution region (1) 
there exists at least one value of ${\rm T_{e,Th}}$ from 
the interval ${\rm (5 - 11)~MeV}$, for which  
the seasonal variation effect in the solar neutrino 
sample of events with recoil electrons 
having ${\rm T_{e} \geq T_{e,Th}}$ is either {\it maximal}
or {\it very close to maximal} (in absolute value).   
Since the precise value of $\Delta m^2$ is still unknown, 
this implies that one can 
effectively search for the seasonal effect 
by forming a set of samples of solar neutrino events, 
which correspond to a sufficiently large 
number of different values of 
${\rm T_{e,Th}}$ from the above interval,
say, for those in Tables I - III,
and by measuring the seasonal variation 
in each of these samples. As we have already 
emphasized, the seasonal effect can change dramatically
with the sample.

 The correlation between the magnitude of the 
seasonal asymmetry and the electron threshold  kinetic energy 
${\rm T_{e,Th}}$ discussed above 
is a manifestation of the fact that 
the underlying statistical distribution of 
neutrino events in the case of interest is a bivariate 
function of time and energy, whose marginal 
distributions are the one-year 
average electron spectrum and the energy integrated time dependence.
An example of a possible graphical representation of 
such a bivariate  distribution of the $(\nu,e^{-})$ events  
is given in Fig. 3a. The figure corresponds to 
$\Delta m^2 = 0.7\times 10^{-10}$\ eV$^2$\ 
and $\sin^22\theta = 0.85$.
Each color represents a specific 
value of the one month average event rate
ratio ${\rm N_{NGF}(t_c,\Delta t = T/12; T_{e,Th})}$, 
eq. (12), calculated for 
${\rm t_c = 0~(perihelion);~ T/12;~ 2T/12;...;
~t_a = T/2~(aphelion)}$. Other types of 
representations are possible, 
of course, but the tests we 
have performed indicate that this one is 
better suited for optimizing the efficiency  
of the information extraction. 
These maps allow to compare at a glance the 
predicted time and energy dependence of the ratio 
${\rm N_{NGF}(t_c,\Delta t; T_{e,Th})}$
for different values of $\Delta m^2$ and $\sin^22\theta$.
For example, Fig. 3a clearly shows that for the chosen 
$\Delta m^2$ and $\sin^22\theta$
no observable time modulation is expected at 
${\rm T_{e,Th}} \cong 5$ MeV,  
while the maximal modulation takes place
for ${\rm T_{e,Th}} \cong 9.5$ MeV. 
When applied to real data, 
the presentation technique used to generate 
Fig. 3a will produce an ``image'' of the bivariate 
event rate distribution, similar to usual 
astronomical images obtained by X-rays or 
$\gamma$-rays telescopes.
If the solar neutrino problem is indeed caused 
by vacuum oscillations of solar neutrinos,
the event rate distribution ``images''
will be characterized 
by a pattern composed of 
elongated {\em spots} of different intensity, 
whose distribution and shape are  functions of the 
neutrino parameters $\Delta m^2$ and $\sin^22\theta$. 
It should be possible to analyze such 
images using well tested techniques, 
as those commonly applied for 
imaging in high energy astronomy 
in order to ensure not 
only the presence of relevant patterns, but also 
their consistency with the 
vacuum oscillations scenario, discarding in 
this way subtle systematical 
errors induced, for instance, by background time variability. 

  As we have shown, for the $^{8}$B neutrino 
induced event rates of interest and 
the VO solution neutrino parameters from 
the region (1) - (2) one has: 
$${\rm N_{NGF}(t_c,\Delta t; T_{e,Th}) 
     = q(T_{e,Th}) + m(T_{e,Th}, \Delta t)~ 
\cos 2\pi {t_c\over T} + O(10^{-3})}~,~\eqno(26)$$

\noindent where 
$q$ is a function of ${\rm T_{e,Th}}$, $m$ depends both on 
${\rm T_{e,Th}}$ and on the sampling interval $\Delta t$,
but both $q$ and $m$ do not depend on $t_c$.
If the data does not exhibit any time dependence, 
apart from that caused by the geometrical effect which is removed from
${\rm N_{NGF}(t_c,\Delta t; T_{e,Th})}$, 
one should have ${\rm m(T_{e,Th}, \Delta t)} = 0$.
The signature of vacuum oscillations is a significant
threshold energy dependent
deviation of ${\rm m(T_{e,Th}, \Delta t)}$ from zero.
For the choice \cite{KP3} of the event rate 
normalization, eq. (12), we have
${\rm q(T_{e,Th})} = 1$ for any sampling scheme (see eq. (15))
and, as it follows from eqs. (14), (15) and (26), 
$${\rm m(T_{e,Th}, \Delta t) = 2\epsilon W(T_{e,Th})}~
{\rm {{\sin2\pi \Delta t/(2T)}\over
{2\pi \Delta t/(2T)}}}~, \eqno(27)$$

\noindent where ${\rm W(T_{e,Th})}$ is given in eq. (19). 
These relations do not necessarily hold true for 
event rate normalizations 
which differ from that used by us. 
Equation (26) allows to reduce the bivariate 
distribution ${\rm N_{NGF}(t_c,\Delta t; T_{e,Th})}$
to the univariate distribution 
${\rm m(T_{e,Th}, \Delta t)}$ without information loss, 
to the extent that eq. (26) represents a good approximation 
to the true time dependence.
In Fig. 3b we compare 
the time dependence of the 
one month average event rates 
${\rm  N_{NGF}(t_c,\Delta t = T/12; T_{e,Th})}$
computed utilizing eqs. (5a), (5b) and (12)
(points with crosses), with 
those obtained using eq. (26),
for ${\rm T_{e,Th}} = 5.0$ MeV (solid line), $7.5$ MeV (dotted line) 
and $10.0$ MeV (dash-dotted line).
As Fig. 3b indicates, eq. (26) describes the time-dependence 
of ${\rm N_{NGF}(t_c,\Delta t = T/12; T_{e,Th})}$
with a very high accuracy. This is confirmed by 
a detailed analysis based on the standard Pearson index,
performed for $\Delta m^2$ in the range 
$(0.1 \div 4.4) \times 10^{-10}$ eV$^2$, 
$\sin^22\theta \leq 1$ and ${\rm T_{e,Th}} < 14.0$ MeV. 
The accuracy of eq. (26) increases quickly  
with the increasing of 
the magnitude of the seasonal asymmetry. 
The largest errors occur for combinations of $\Delta m^2$, 
$\sin^22\theta$ and ${\rm T_{e,Th}}$ 
for which the predicted NGF asymmetry is negligible, 
i.e., is smaller than 0.6\%. 
It is not surprising that eq. (26) is
very accurate even for 
$\Delta m^2$ as large as 
$4.4 \times 10^{-10}$\ eV$^2$
given the results obtained 
at the end of the previous Section (see eqs. (20) - (22) and
the related discussion).

  As we have already  indicated, the function 
${\rm W(T_{e,Th})}$ 
carries all the information about 
the threshold energy dependence of 
the event rates and the seasonal asymmetries 
of interest, eqs. (15) - (18);
its knowledge allows to reconstruct 
the time dependence of
the normalized event rates (12).
Figure 3c and Table III illustrate 
the dependence of ${\rm W}$ on 
${\rm T_{e,Th}}$ for the representative
set of values of the neutrino 
parameters $\Delta m^2$ and $\sin^22\theta$. 
The function ${\rm W(T_{e,Th})}$ was calculated 
numerically by fitting the 
time-dependence of ${\rm N_{NGF}(t_c,T/12; T_{e,Th})}$
(computed without any approximations) 
with that of eq. (26), 
utilizing the least squares method.
The maximal value of ${\rm W(T_{e,Th})}$
for each pair of $\Delta m^2$ and $\sin^22\theta$,
${\rm W^{max}}$, and the value of
${\rm T_{e,Th}}$ at which it takes place, 
${\rm T^{max}_{e,Th}}$, are also given in  Table III.

    Let us consider an alternative 
event rate normalization and the corresponding 
version of eq. (26), i.e.,
${\rm R_{NGF}(t_c, \Delta t; T_{e,Th})/R_{SSM}(T_{e,Th}) 
  = q'(T_{e,Th}) + m'(T_{e,Th}, \Delta t)~cos 2\pi t_c/T}$,
where ${\rm q'(T_{e,Th})} \neq 1$ represents 
the ``distortion'' of the one year average event 
rate distribution and 
${\rm R_{SSM}(T_{e,Th})}$ is the one year average 
event rate predicted in a standard solar model
in the absence of neutrino oscillations. 
Now ${\rm m'(T_{e,Th}, \Delta t)}$  
carries  information about 
the amplitude of the time variation, while
both ${\rm m'(T_{e,Th}, \Delta t)}$ and
${\rm q'(T_{e,Th})}$ determine  
the energy dependence.
However, since they are related to the 
index ${\rm m(T_{e,Th}, \Delta t)}$ in eq. (26)  by
${\rm m(T_{e,Th}, \Delta t) = 
m'(T_{e,Th}, \Delta t)/q'(T_{e,Th}, \Delta t)}$,
it seems more convenient to use the 
normalization in eq. (12) leading 
to eq. (26) with only one function 
of ${\rm T_{e,Th}}$
(${\rm q(T_{e,Th}) = 1}$),
rather than having the
${\rm T_{e,Th}}$ dependence spread in 
two functions. 

   If the geometrical effect is included 
in the event rate data,
it can be accounted for in the observables
(12) and (13), as it follows from eqs. (16) and (18), 
by simply adding $1$ 
to the function  
${\rm W(T_{e,Th})}$ (e.g., in eq. (27)). 
This is a consequence of the normalization \cite{KP3}
used by us in eq. (12). There exist also 
simple relations between the function
${\rm W(T_{e,Th})}$ and the different 
types of seasonal asymmetries considered by other authors. 
For instance, relation (17) holds for the asymmetry 
${\rm A'_{NGF}(\Delta t; T_{e,Th}) 
= 2(R_{NGF}(0,\Delta t; T_{e,Th}) - 
R_{NGF}(T/2,\Delta t; T_{e,Th})/
(R_{NGF}(0,\Delta t; T_{e,Th})}$ + 
${\rm R_{NGF}(T/2,\Delta t; T_{e,Th})}$. 
The asymmetry defined in \cite{MSVO98} 
corresponds to 
${\rm A_S =}$ 2${\rm (R_{GF}(t_c = 0) - R_{GF}(t_c = T/2)/(R_{GF}(t_c = T/4) 
+ R_{GF}(t_c = 3T/4))}$ with ${\rm \Delta t = T/4}$ and
${\rm T_{e,Th} = 6.5~MeV}$ and
is given by ${\rm A_S} =$ 
4${\rm \epsilon {{2\sqrt{2}}\over\pi} (1 + W(6.5~MeV))}$. 
Evidently, eq. (26) can be used to relate 
different seasonal asymmetries 
and spectral distortion functions 
discussed in the literature, while 
plots and tables like Fig. 3c and Table III can be utilized 
to make predictions for the latter for a large range of sampling 
schemes, ${\rm T_{e,Th}}$ and values of the neutrino 
oscillation parameters.

\vskip 0.3cm
\leftline{\bf 4. Conclusions}
\vskip 0.3cm
\indent We have studied in detail the 
threshold energy (${\rm T_{e,Th}}$) dependence
of the seasonal variation effect
in the electron energy 
integrated solar neutrino signal 
of the Super-Kamiokande
detector in the case of the vacuum oscillation
solution of the solar neutrino problem. 
We have shown that for the values of 
the neutrino oscillation parameters
$\Delta m^2$ and $\sin^22\theta$
from the VO solution region (1) - (2),
the time and the ${\rm T_{e,Th}}$
dependence of the corresponding 
solar neutrino event rate
factorize to a high degree of accuracy.
This facorization takes a particularly simple form, 
eqs. (15) - (18),
when the event rate for a given
threshold energy ${\rm T_{e,Th}}$ is normalized to the one 
year average event rate for the same
threshold energy \cite{KP3}, eq. (12).
As a consequence of the factorization, the VO induced seasonal 
variation asymmetry is proportional to the geometrical one,
the coefficient of proportionality being a  
function of ${\rm T_{e,Th}}$, but not of time.
The asymmetry which contains both the 
VO induced and the geometrical seasonal effects,
${\rm A^{seas}_{GF}}$, is just equal to the sum 
of the seasonal asymmetry due to the VO only,
${\rm A^{seas}_{NGF}}$, and of
the geometrical asymmetry ${\rm A^{seas}_{geom}}$ (eq. (18)).
The asymmetries
${\rm A^{seas}_{NGF}}$ and ${\rm A^{seas}_{geom}}$
can mutually cancel and ${\rm A^{seas}_{GF}}$
can be close to zero \cite{KP3}.
This implies, in particular, that one cannot have
simultaneously 
$|{\rm A^{seas}_{NGF}| \ll A^{seas}_{geom}}$ and 
$|{\rm A^{seas}_{GF}| \ll A^{seas}_{geom}}$
for the same set of values of the parameters. However,
one of the two asymmetries, 
${\rm A^{seas}_{NGF}}$ or ${\rm A^{seas}_{GF}}$, 
can be strongly suppressed, while the other can have 
observable values. All these 
possibilities are realized for the values of
$\Delta m^2$, $\sin^22\theta$, 
${\rm E_{\nu}}$ and ${\rm T_{e,Th}}$ 
of interest.

   The seasonal variation effect 
exhibits a strong dependence on 
${\rm T_{e,Th}}$. For the values 
${\rm \Delta m^2 \cong (4.3 - 4.4)\times 10^{-10}~eV^2}$ 
and $\sin^22\theta \cong 0.9$, suggested by the 
current Super-Kamiokande data 
on the e$^{-}-$spectrum \cite{SKSmyDPF99}, for instance,
the VO induced seasonal asymmetry 
is negligible for ${\rm T_{e,Th} \cong (5 - 8)~MeV}$, 
increases with ${\rm T_{e,Th}}$ reaching a maximum at 
${\rm T_{e,Th} \cong 10.1~MeV}$, 
${\rm A^{seas}_{NGF}(T_{e,Th} = 10.1~MeV)} \cong  14.8\% $,
then decreases to zero at
${\rm T_{e,Th} \cong 12.2~MeV}$ and becomes negative for
larger values of ${\rm T_{e,Th}}$.
The geometrical effect amplifies the asymmetry 
in the region of the maximum and 
${\rm A^{seas}_{GF}(T_{e,Th} = 10.1~MeV)} \cong  21.5\% $
(Fig. 2c). At ${\rm T_{e,Th} \cong 9.5~(11.0)~MeV}$, we have 
for the indicated values of ${\rm \Delta m^2}$ and $\sin^22\theta$:
${\rm A^{seas}_{NGF}} \cong  13.0~(10.0)\% $
\footnote{\tightenlines \footnotesize 
The quoted asymmetry values correspond to ideal
experimental conditions; they will be somewhat 
smaller, by $\sim (1 - 2)\%$, if one takes into account the
e$^{-}$ detection efficiency and energy resolution 
of the Super-Kamiokande detector}. Results for other 
VO solution values of ${\rm \Delta m^2}$ 
and $\sin^22\theta$ can be found in Table II.

  We have shown also that 
for any given $\Delta m^2$ and $\sin^22\theta$ 
from the VO solution region (1) - (2)
there exists at least one value of ${\rm T_{e,Th}}$ from 
the interval ${\rm (5 - 11)~MeV}$, for which  
the VO induced seasonal 
effect in the solar neutrino 
sample of events with recoil electrons 
having ${\rm T_{e} \geq T_{e,Th}}$ is either {\it maximal}
or {\it very close to maximal}.   
Given the fact that the precise values of $\Delta m^2$ 
and $\sin^22\theta$ are still unknown, 
this suggests that one can 
effectively search for the seasonal effect 
by forming a set of samples of solar neutrino events, 
which correspond to a sufficiently large 
number of different values of 
${\rm T_{e,Th}}$ from the above interval,
say, for those in Tables I - III,
${\rm T_{e,Th} =}$ 5; 6; 7;...; 11 MeV,
and by measuring the seasonal variation 
in each of these samples. The seasonal 
effect can change dramatically
with the sample. The results summarized above 
are illustrated in Figs. 1 - 3 and Tables II - III.
Although they have been derived 
for the Super-Kamiokande
detector, we expect similar results 
to be valid for the SNO and ICARUS detectors. 

\vskip 0.3cm
\leftline{\bf Acknowledgements.} 
The work of S.T.P. was supported in part by the Italian MURST
under the program ``Fisica Teorica delle
Interazioni Fondamentali''  and by Grant PH-510 from the
Bulgarian Science Foundation.

\newpage
\tightenlines

\newpage

\def\solidline{\mbox{$^{\_\!\_\!\_\!\_\!\_\!\_}$}}
\def\dottedline{\mbox{$^{.......}$}}
\def\dashdottedline{\mbox{$^{\_.\_.\_}$}}
\def\dashedline{\mbox{$^{\_\,\,\_\,\,\_}$}}
\def\squaresline{\mbox{{\small $\Box$}}}
\def\diamondsline{\mbox{{\small $\Diamond$}}}
\def\starsline{\mbox{$\star$}}
\def\trianglesline{\mbox{$\triangle$}}
\tightenlines
\renewcommand{\arraystretch}{0.5}
\begin{table} 
\noindent \caption{The ratio 
${\rm R(T_{e,Th})/R(5~MeV)}$
of the one year average event rates 
due to $^{8}$B neutrinos in  
$\nu - e^{-}$ 
detectors (Super-Kamiokande, etc.) 
for several values of ${\rm T_{e,Th}}$ 
and for a set of values of $\Delta m^2$ and $\sin^22\theta$ 
from the VO solution region.
}
\vskip -0.6truecm
\begin{center}
 \begin{tabular}{|cc||cc||rrrrrrr|}
\hline
\multicolumn{2}{|c||}{n} & $\Delta m^2$        &          &
\multicolumn{7}{c|}{$T_{e,Th}$~(MeV)}\\
 & & ($10^{-10}$ eV$^2$) & $\sin^2 2\theta$  &  
\multicolumn{1}{c}{5.5} &  
\multicolumn{1}{c}{6.0} &
\multicolumn{1}{c}{7.0} &
\multicolumn{1}{c}{8.0} &
\multicolumn{1}{c}{9.0} &
\multicolumn{1}{c}{10.0} & 
\multicolumn{1}{c|}{11.0} \\
\hline
\hline
\multicolumn{4}{|c||}{no oscillations}     & 0.84 & 0.69 & 0.44 & 0.26&
0.14 & 0.06 & 0.02 \\ \hline \hline
1 & \solidline\ &0.40 & 1.00 & 0.86 & 0.73 & 0.50 & 0.31 & 0.17 & 0.08 & 0.03
\\ 
\hline
2 & \dottedline\ &0.50 & 0.60 & 0.85 & 0.71 & 0.47 & 0.29 & 0.16 & 0.07 & 0.03\\
3 & \dashdottedline\ &0.50 & 0.90 & 0.86 & 0.73 & 0.50 & 0.32 & 0.18 & 0.09 &
0.03 \\ \hline
4 & \dashedline\ &0.55 & 1.00 & 0.87 & 0.74 & 0.52 & 0.34 & 0.20 & 0.10 & 0.04
\\ \hline
5 & \squaresline\ &0.60 & 0.70 & 0.84 & 0.70 & 0.47 & 0.29 & 0.16 & 0.08 &
0.03 \\
6 & \diamondsline\ &0.60 & 0.90 & 0.85 & 0.72 & 0.49 & 0.31 & 0.18 & 0.09 &
0.03 \\ \hline
7 & \starsline\ &0.65 & 0.85 & 0.84 & 0.70 & 0.47 & 0.29 & 0.17 & 0.08 & 0.03
\\ \hline
8 & \trianglesline\ &0.70 & 0.85 & 0.82 & 0.67 & 0.44 & 0.27 & 0.16 & 0.08 &
0.03 \\ \hline
9 & \solidline\ &0.75 & 0.80 & 0.81 & 0.65 & 0.42 & 0.25 & 0.14 & 0.07 & 0.03
\\ \hline
10 & \dottedline\ &0.80 & 0.65 & 0.81 & 0.65 & 0.41 & 0.24 & 0.13 & 0.06
& 0.02 \\
11 &\dashdottedline\ &0.80 & 0.70 & 0.81 & 0.65 & 0.41 & 0.24 & 0.13 & 0.06 &
0.02 \\
12 & \dashedline\ &0.80 & 0.85 & 0.79 & 0.63 & 0.38 & 0.23 & 0.12 & 0.06 &
0.02 \\ \hline
13 & \squaresline\ &0.85 & 0.75 & 0.80 & 0.63 & 0.38 & 0.22 & 0.12 & 0.05 &
0.02 \\
14 & \diamondsline\ &0.85 & 1.00 & 0.75 & 0.56 & 0.31 & 0.17 & 0.09 & 0.04 &
0.02 \\ \hline
15 & \starsline\ &0.90 & 0.75 & 0.79 & 0.62 & 0.37 & 0.21 & 0.11 & 0.05 & 0.02
\\
16 & \trianglesline\ &0.90 & 1.00 & 0.75 & 0.55 & 0.29 & 0.15 & 0.07 & 0.03 &
0.01\\ 
\hline
17 & \solidline\ &1.00 & 0.75 & 0.79 & 0.62 & 0.36 & 0.19 & 0.09 & 0.04 & 0.02
\\
18 & \dottedline\ &1.00 & 1.00 & 0.76 & 0.56 & 0.29 & 0.13 & 0.06 & 0.02
& 0.01 \\  \hline
19 & \dashdottedline\ &1.10 & 1.00 & 0.78 & 0.60 & 0.31 & 0.14 & 0.06 & 0.02 &
0.01 \\ \hline
20 & \dashedline\ &1.20 & 1.00 & 0.81 & 0.63 & 0.35 & 0.17 & 0.07 & 0.03 &
0.01 \\ \hline
21 & \squaresline\ &1.30 & 1.00 & 0.83 & 0.67 & 0.39 & 0.20 & 0.09 & 0.03 &
0.01 \\ \hline
22 & \diamondsline\ &2.50 & 0.90 & 0.83 & 0.67 & 0.39 & 0.22 & 0.12 & 0.07 &
0.03 \\ \hline
23 & \starsline\ &4.40 & 0.90 & 0.83 & 0.68 & 0.44 & 0.25 & 0.13 & 0.07 & 0.03
\\ \hline
\end{tabular}
\end{center}
\end{table}

\vskip -0.6truecm
\renewcommand{\arraystretch}{0.5}
\begin{table} 
\noindent \caption{
The perihelion - aphelion 
asymmetry ${\rm A_{NGF}^{seas}(t_c = 0,\Delta t =T/12; T_{e,Th}}$) 
$\equiv {\rm A_{NGF}^{seas}(T_{e,Th})}$ in \% 
for different values of ${\rm T_{e,Th}}$ and a set of
values of $\Delta m^2$ and $\sin^22\theta$ 
from the VO solution region.
Given also are the values of the asymmetry corresponding 
to a maximum of $|{\rm A_{NGF}^{seas}}|$ in the interval
${\rm T_{e,Th} = (5 - 11)~MeV}$, ${\rm A_{NGF}^{seas,max}}$, 
and the ${\rm T_{e,Th}}$ at which it is reached, ${\rm T^{max}_{e,Th}}$. 
The asymmetry 
including the geometrical effect
${\rm A_{GF}^{seas}(0,T/12; T_{e,Th}}$) 
$ = {\rm A_{NGF}^{seas}(T_{e,Th})} + 4\epsilon$.
}
\vskip -0.6truecm
\begin{center}
\begin{tabular}{|cc||rr||rrrrrrr|}
\hline
$\Delta m^2$ &   
& 
\multicolumn{1}{c}{${\rm T_{e,Th}^{max}}$} &   
   & 
\multicolumn{7}{c|}{${\rm T_{e,Th}}$\ (MeV)}
\\
$(10^{-10}$ eV$^2)$ & $\sin^2 2\theta$ &  
\multicolumn{1}{c}{(MeV)}      & 
\multicolumn{1}{c||}{${\rm A_{NGF}^{seas, max}}$} &  
\multicolumn{1}{c}{5.0} &
\multicolumn{1}{c}{6.0} &
\multicolumn{1}{c}{7.0} &
\multicolumn{1}{c}{8.0} &
\multicolumn{1}{c}{9.0} &
\multicolumn{1}{c}{10.0} &
\multicolumn{1}{c|}{11.0} \\
\hline
\hline
0.40 & 1.00 & 5.00 & 4.05 & 4.05 & 3.73 & 3.38 & 3.04 & 2.73 & 2.45 & 2.20 \\
\hline
0.50 & 0.60 & 6.28 & 2.16 & 2.08 & 2.16 & 2.14 & 2.05 & 1.94 & 1.81 & 1.68 \\
0.50 & 0.90 & 5.66 & 4.46 & 4.41 & 4.43 & 4.24 & 3.94 & 3.62 & 3.29 & 2.99 \\
\hline
0.55 & 1.00 & 6.07 & 6.39 & 6.02 & 6.39 & 6.24 & 5.83 & 5.34 & 4.83 & 4.36 \\
\hline
0.60 & 0.70 & 7.98 & 3.10 & 2.22 & 2.75 & 3.02 & 3.10 & 3.04 & 2.91 & 2.75 \\
0.60 & 0.90 & 7.45 & 5.25 & 4.03 & 4.90 & 5.23 & 5.19 & 4.95 & 4.61 & 4.25 \\
\hline
0.65 & 0.85 & 8.54 & 4.82 & 2.46 & 3.71 & 4.48 & 4.79 & 4.79 & 4.61 & 4.34 \\
\hline
0.70 & 0.85 & 9.46 & 5.00 & 0.97 & 2.63 & 3.92 & 4.67 & 4.97 & 4.97 & 4.79 \\
\hline
0.75 & 0.80 & 10.85 & 4.43 & -0.68 & 0.86 & 2.33 & 3.43 & 4.08 & 4.38 & 4.42 \\
\hline
0.80 & 0.65 & 11.00 & 2.82 & -1.38 & -0.47 & 0.56 & 1.46 & 2.14 & 2.58 & 2.82 \\
0.80 & 0.70 & 11.00 & 3.28 & -1.60 & -0.55 & 0.66 & 1.72 & 2.51 & 3.01 & 3.28 \\
0.80 & 0.85 & 11.00 & 5.19 & -2.56 & -0.91 & 1.11 & 2.92 & 4.18 & 4.90 & 5.19 \\
\hline
0.85 & 0.75 & 11.00 & 3.54 & -3.01 & -2.06 & -0.64 & 0.84 & 2.08 & 2.99 & 3.54 \\
0.85 & 1.00 & 11.00 & 8.99 & -6.73 & -5.18 & -1.78 & 2.46 & 6.00 & 8.15 & 8.99 \\
\hline
0.90 & 0.75 & 5.00 & -3.89 & -3.89 & -3.33 & -2.11 & -0.58 & 0.90 & 2.13 & 2.99 \\
0.90 & 1.00 & 5.29 & -8.22 & -8.12 & -7.85 & -5.64 & -1.70 & 2.74 & 6.31 & 8.35 \\
\hline
1.00 & 0.75 & 5.86 & -5.07 & -4.75 & -5.06 & -4.61 & -3.49 & -1.99 & -0.39 & 1.03 \\
1.00 & 1.00 & 6.72 & -10.56 & -8.68 & -10.17 & -10.49 & -9.13 & -5.85 & -1.22 & 3.29 \\
\hline
1.10 & 1.00 & 8.25 & -12.24 & -7.61 & -9.77 & -11.45 & -12.23 & -11.66 & -9.33 & -5.36 \\
\hline
1.20 & 1.00 & 9.46 & -13.40 & -5.88 & -8.18 & -10.39 & -12.15 & -13.21 & -13.24 & -11.83 \\
\hline
1.30 & 1.00 & 10.85 & -14.11 & -3.98 & -6.15 & -8.53 & -10.69 & -12.45 & -13.68 & -14.09 \\
\hline
2.50 & 0.90 & 9.79 & 14.31 & 0.10 & -0.12 & 0.87 & 7.10 & 13.18 & 14.01 & 11.44 \\
\hline
4.40 & 0.90 & 10.13 & 14.75 & 0.92 & 1.04 & 2.34 & 1.62 & 8.59 & 14.69 & 9.99 \\
\hline
\end{tabular}
\end{center}
\end{table}
\vskip -0.6cm
\renewcommand{\arraystretch}{0.5}
\begin{table}\label{tab:tab3a}
\caption{Values of the function ${\rm W(T_{e,Th})}$
for various ${\rm T_{e,Th}}$ and 
$\Delta m^2$  and $\sin^22\theta$.}
\begin{center}
\begin{tabular}{|cc||rr||rrrrrrr|}
\hline
$\Delta m^2$         &
                     &
\multicolumn{1}{c}{ ${\rm T_{e,Th}^{max}}$ }&
          &
\multicolumn{7}{c|}{${\rm T_{e,Th}}$\ (MeV)}
\\
($10^{-10}$\ eV$^2$) & 
$\sin^2 2\theta$     &
\multicolumn{1}{c}{    (MeV)}        & 
\multicolumn{1}{c||}{${\rm W^{max}}$} &
\multicolumn{1}{c}{5.0} &
\multicolumn{1}{c}{6.0} &
\multicolumn{1}{c}{7.0} &
\multicolumn{1}{c}{8.0} &
\multicolumn{1}{c}{9.0} &
\multicolumn{1}{c}{10.0} &
\multicolumn{1}{c|}{11.0} \\
\hline
\hline
0.40 & 1.00 & 5.00 & 0.61 & 0.61 & 0.56 & 0.51 & 0.46 & 0.41 & 0.37 & 0.33 \\
\hline
0.50 & 0.60 & 6.17 & 0.33 & 0.32 & 0.33 & 0.32 & 0.31 & 0.29 & 0.27 & 0.25 \\
0.50 & 0.90 & 5.51 & 0.68 & 0.67 & 0.67 & 0.64 & 0.60 & 0.55 & 0.50 & 0.45 \\
\hline
0.55 & 1.00 & 6.17 & 0.97 & 0.91 & 0.97 & 0.95 & 0.88 & 0.81 & 0.73 & 0.66 \\
\hline
0.60 & 0.70 & 8.02 & 0.47 & 0.34 & 0.42 & 0.46 & 0.47 & 0.46 & 0.44 & 0.42 \\
0.60 & 0.90 & 7.30 & 0.80 & 0.61 & 0.74 & 0.79 & 0.79 & 0.75 & 0.70 & 0.64 \\
\hline
0.65 & 0.85 & 8.49 & 0.73 & 0.38 & 0.56 & 0.68 & 0.73 & 0.73 & 0.70 & 0.66 \\
\hline
0.70 & 0.85 & 9.50 & 0.76 & 0.15 & 0.40 & 0.60 & 0.71 & 0.75 & 0.75 & 0.73 \\
\hline
0.75 & 0.80 & 10.63 & 0.67 & -0.10 & 0.13 & 0.36 & 0.52 & 0.62 & 0.67 & 0.67 \\
\hline
0.80 & 0.65 & 11.00 & 0.43 & -0.21 & -0.07 & 0.09 & 0.22 & 0.33 & 0.39 & 0.43 \\
0.80 & 0.70 & 11.00 & 0.50 & -0.24 & -0.08 & 0.10 & 0.26 & 0.38 & 0.46 & 0.50 \\
0.80 & 0.85 & 11.00 & 0.79 & -0.39 & -0.13 & 0.17 & 0.45 & 0.64 & 0.75 & 0.79 \\
\hline
0.85 & 0.75 & 11.00 & 0.54 & -0.45 & -0.31 & -0.09 & 0.13 & 0.32 & 0.46 & 0.54 \\
0.85 & 1.00 & 11.00 & 1.37 & -1.02 & -0.78 & -0.26 & 0.38 & 0.92 & 1.25 & 1.37 \\
\hline
0.90 & 0.75 & 5.02 & -0.59 & -0.59 & -0.50 & -0.32 & -0.08 & 0.14 & 0.33 & 0.46 \\
0.90 & 1.00 & 5.41 & -1.24 & -1.23 & -1.19 & -0.85 & -0.24 & 0.43 & 0.97 & 1.28 \\
\hline
1.00 & 0.75 & 5.94 & -0.77 & -0.72 & -0.77 & -0.70 & -0.53 & -0.30 & -0.05 & 0.16 \\
1.00 & 1.00 & 6.78 & -1.60 & -1.32 & -1.54 & -1.59 & -1.38 & -0.87 & -0.17 & 0.52 \\
\hline
1.10 & 1.00 & 8.17 & -1.85 & -1.15 & -1.48 & -1.73 & -1.85 & -1.76 & -1.41 & -0.79 \\
\hline
1.20 & 1.00 & 9.50 & -2.03 & -0.90 & -1.24 & -1.58 & -1.84 & -2.00 & -2.00 & -1.78 \\
\hline
1.30 & 1.00 & 10.83 & -2.14 & -0.61 & -0.94 & -1.30 & -1.62 & -1.89 & -2.07 & -2.13 \\
\hline
2.50 & 0.90 & 9.68 & 2.19 & 0.02 & -0.02 & 0.14 & 1.10 & 2.04 & 2.14 & 1.73 \\
\hline
4.40 & 0.90 & 10.05 & 2.30 & 0.14 & 0.16 & 0.37 & 0.25 & 1.35 & 2.29 & 1.50 \\
\hline
\end{tabular}
\end{center}
\end{table}

\renewcommand{\arraystretch}{1.0}

\newpage

\centerline{\bf FIGURE CAPTIONS}

\vskip 0.5cm
\baselineskip 18pt
\noindent {\bf Figure 1.} The ${\rm t_c}$ and ${\rm T_{e,Th}}$ 
dependence of the normalized one month average event rate 
${\rm N_{NGF}(t_c, \Delta t = T/12; T_{e,Th})}$, eq. (12),
in $\nu - e^{-}$ detectors (Super-Kamiokande, etc.)
for six representative values of
$\Delta m^2$ and $\sin^22\theta$ from the VO solution region,
given in Table 1 under the numbers:
n = 3 (a), 8 (b), 16 (c), 21 (d), 22 (e), 23 (f).
The solid, dotted, dash-dotted, dashed lines and 
the lines with squares, diamonds, stars  
in each of the sub-figures (a) - (f)
correspond to seven equally spaced values of 
${\rm t_c} = 0$ (perihelion); T/12; T/6; T/4; T/3; 5T/12; T/2,
respectively.
The seasonal asymmetry is given by
${\rm A^{seas}_{NGF}(t_c, T/12; T_{e,Th})}$ = 
${\rm N_{NGF}(t_c, T/12; T_{e,Th})}$ -
${\rm N_{NGF}(T/2 -t_c, T/12; T_{e,Th})}$,
${\rm 0 \leq t_c \leq T/4}$, and is maximal for ${\rm t_c = 0}$.

\vskip 0.5cm
\noindent {\bf Figure 2.} The dependence of 
the peri\-he\-lion - aphe\-lion asymmetry
${\rm A^{seas}_{NGF}(t_c = 0,\Delta t = T/12}$; ${\rm T_{e,Th}})$ 
(in \%), eq. (13), 
on ${\rm T_{e,Th}}$ for different values of $\Delta m^2$
and $\sin^22\theta$ from the VO solution region.
The curves in the panels (a), (b) and (c) correspond 
to the values of $\Delta m^2$ and 
$\sin^22\theta$ in Table I  
numbered, respectively: 1 - 8, 9 - 16 and 17 - 23.
Each curve of a given type (solid, dashed, dash-dotted, etc.)
is obtained for the values of $\Delta m^2$ and 
$\sin^22\theta$ which are marked with the symbol 
of the curve
(solid line, dashed line, dash-dotted line etc.)
in the second column of Table 1.  
The asymmetry which includes the geometrical effect
${\rm A^{seas}_{GF}(0, T/12; T_{e,Th})}$
= ${\rm A^{seas}_{NGF}(0, T/12; T_{e,Th})} + 4\epsilon$,
$4\epsilon = 6.68\% $, is also shown.

\vskip 0.5cm
\noindent {\bf Figure 3.} a). Am ``image'' of the one month 
average event rate ratio (distribution)  
${\rm N_{NGF}(t_c, T/12}$;${\rm T_{e,Th})}$ (eq. (12)). 
Each vertical stripe 
corresponds to ${\rm t_c = 0;~T/12;~T/6;~T/4;...;~T}$, 
as is indicated on the horizontal axis.
The vertical scale on the right-hand side 
allows to convert a color into a 
value of the ratio. 
b) The event rate ratio shown in a), plotted as a function
of ${\rm \cos 2\pi t_c/T}$ for 
${\rm T_{e,Th}} = 5.0$ MeV (solid line), 
        $7.5$ MeV (dotted line),
       $10.0$ MeV (dashed-dotted line). The 
lines are the best fits to the computed ratio indicated by ``$+$''.
 c) The dependence of ${\rm W(T_{e,Th})}$
on ${\rm T_{e,Th}}$. All results shown in the 
figure are derived for $\Delta m^2 = 0.70 \times 10^{-10}$\ eV$^2$ and
$\sin^22\theta = 0.85$.


\baselineskip 18pt
\begin{thebibliography}{99}

\bibitem{Pont67} B. Pontecorvo, Zh. Eksp. Teor. Fiz. 53 (1967) 1717;
V. Gibov and B. Pontecorvo, Phys. Lett. B28 (1969) 493; 
S.M. Bilenky and B. Pontecorvo, Phys. Rep. 41 (1978) 225.

\bibitem{Pont57} B. Pontecorvo, Zh. Eksp. Teor. Fiz. 33 (1957) 549; ibid.
34 (1958) 247; Z. Maki, M. Nakagawa and S. Sakata,
Prog. Theor. Phys. 28 (1962) 870.

\bibitem{BPet87} S.M. Bilenky and S.T. Petcov, 
Rev. Mod. Phys. {\bf 59} (1987) 671.

\bibitem{KP1} P.I. Krastev and S.T. Petcov, Phys. Lett. 285B (1992) 85;
ibid. 299B (1993) 99; 
V. Barger et al., Phys. Rev. Lett. 69 (1992) 3135, and
Phys. Rev. D43 (1991) 1110; A. Acker, S. Pakvasa and
J. Pantaleone, Phys. Rev. Lett. 65 (1990) 2479, and
Phys. Rev. D43 (1991) 1754; see also: S.L. Glashow and L.M. Krauss, Phys.
Lett. 190B (1987) 199.

\bibitem{KP2} P.I. Krastev and S.T. Petcov, Phys. Rev. Lett. 72 (1994) 1960.
\bibitem{KP3}  P.I. Krastev and S.T. Petcov, Nucl. Phys. B449 (1996) 605.
\bibitem{KP4}  P.I. Krastev and S.T. Petcov, Phys. Rev. D53 (1996) 1665.

\bibitem{SKSmyDPF99} M. Smy, Super-Kamiokande Coll.,
Talk given at the DPF 
Conference, American Institute of Physics, January 5 - 7, 1999, Los Angeles
(see the WWW page of the Conference: 
www.physics.ucla.edu/dpf99/trans/DPF99\_Transparencies.htm).

\bibitem{BKS98} J.N. Bahcall, P.I. Krastev and A.Yu. Smirnov, 
Phys. Rev. D58 (1998) 096016.

\bibitem{Cl} R. Davis et al., Prog. Part. Nucl. Phys. 32 (1994) 13;
K. Lande, Homestake Coll., Talk given at the {\it Neutrino '98} International
Conference \cite{nu98}, 1998.

\bibitem{Kam} K.S. Hirata et al., Kamiokande Coll., 
Phys. Rev. Lett. 77 (1996) 1683.

\bibitem{GALLEX} W. Hampel et al., GALLEX Coll.,  Phys. Lett. B388
(1996) 384; T. Kirsten, GALLEX Coll., 
Talk given at the {\it Neutrino '98} International
Conference \cite{nu98}, 1998.

\bibitem{SAGE} D.N. Abdurashitov et al., SAGE Coll., Phys. Rev. Lett. 
77 (1996) 4708; V. Gavrin, SAGE Coll.,
Talk given at the {\it Neutrino '98} International
Conference \cite{nu98}, 1998.

\bibitem{SKSuzukinu98} Y. Suzuki, Super-Kamiokande Coll.,
Talk given at the {\it Neutrino '98} International
Conference \cite{nu98}, 1998;
Y. Fukuda et al., Phys. Rev. Lett. 81 (1998) 1158.

\bibitem{JNB}
J.N. Bahcall, {\it Neutrino Astrophysics}, Cambridge University
Press, Cambridge, 1989.

\bibitem{BP98}
J.N. Bahcall, S. Basu and M. Pinsonneault, Phys. Lett. B433 (1998) 1.

\bibitem{FVO98} B. Faid et al., Phys. Rev. D56 (1997) 4374; hep-ph/9805293.

\bibitem{MSVO98} S.P. Mikheyev and A.Yu. Smirnov, 
Phys. Lett. B429 (1998) 343.

\bibitem{GlVO98} S.L. Glashow,
P.J. Kernan and L. Krauss, hep-ph/9808470.

\bibitem{BeVO98} V. Berezinsky, 
G. Fiorentini and M. Lissia, hep-ph/9811352. 

\bibitem{BarVO98} V. Barger and K. Whisnant, hep-ph/9812273.

\bibitem{SPJR89} S.T. Petcov and J. Rich, Phys. Lett. B224 (1989) 401.

\bibitem{nu98} International Conference on 
Neutrino Physics and Astrophysics
{\it Neutrino '98}, June 3 - 9, 1998, Takayama, Japan;
WWW page: http://www-sk.icrr.u-tokyo.ac.jp/nu98.

\end{thebibliography}
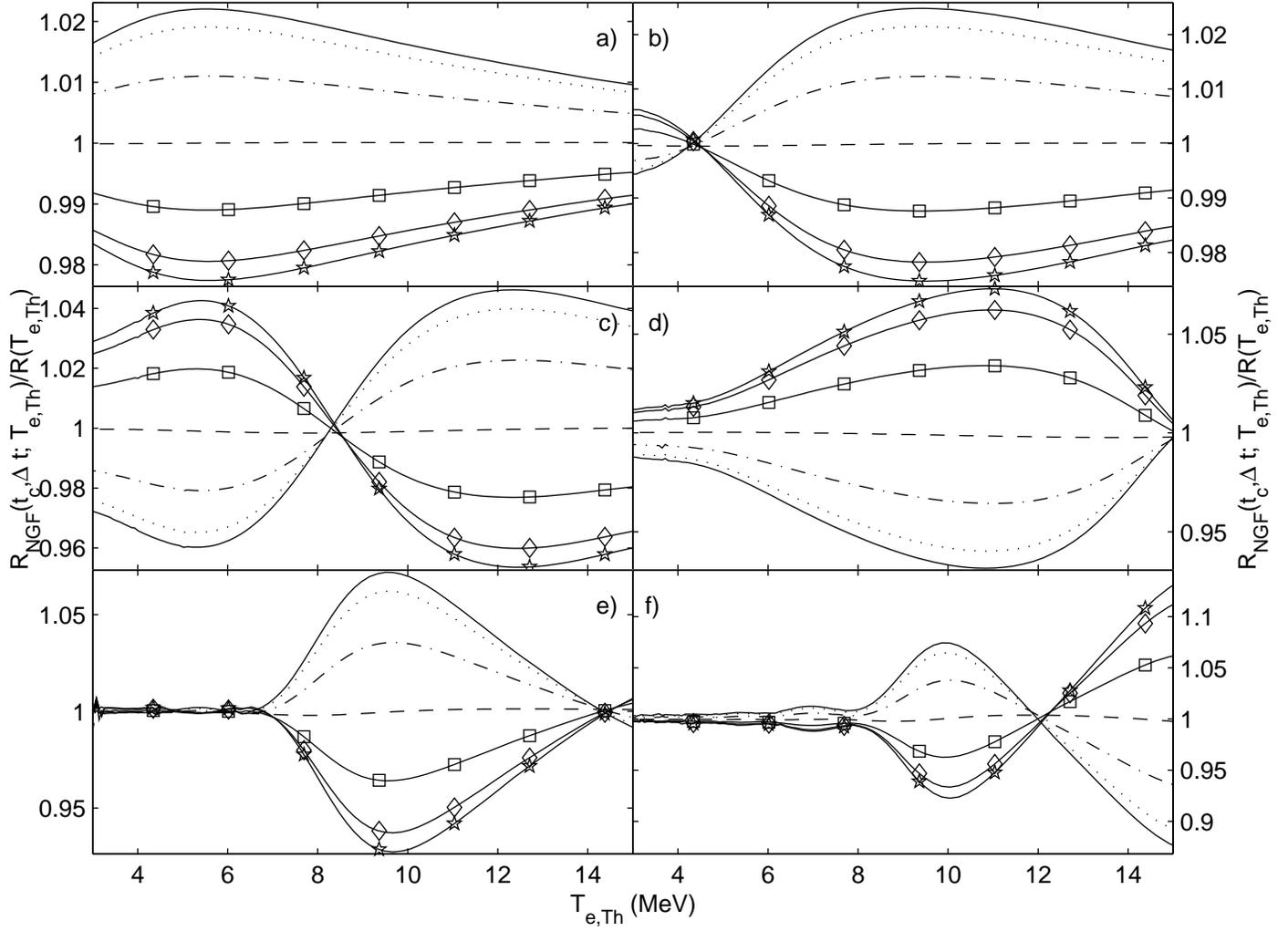
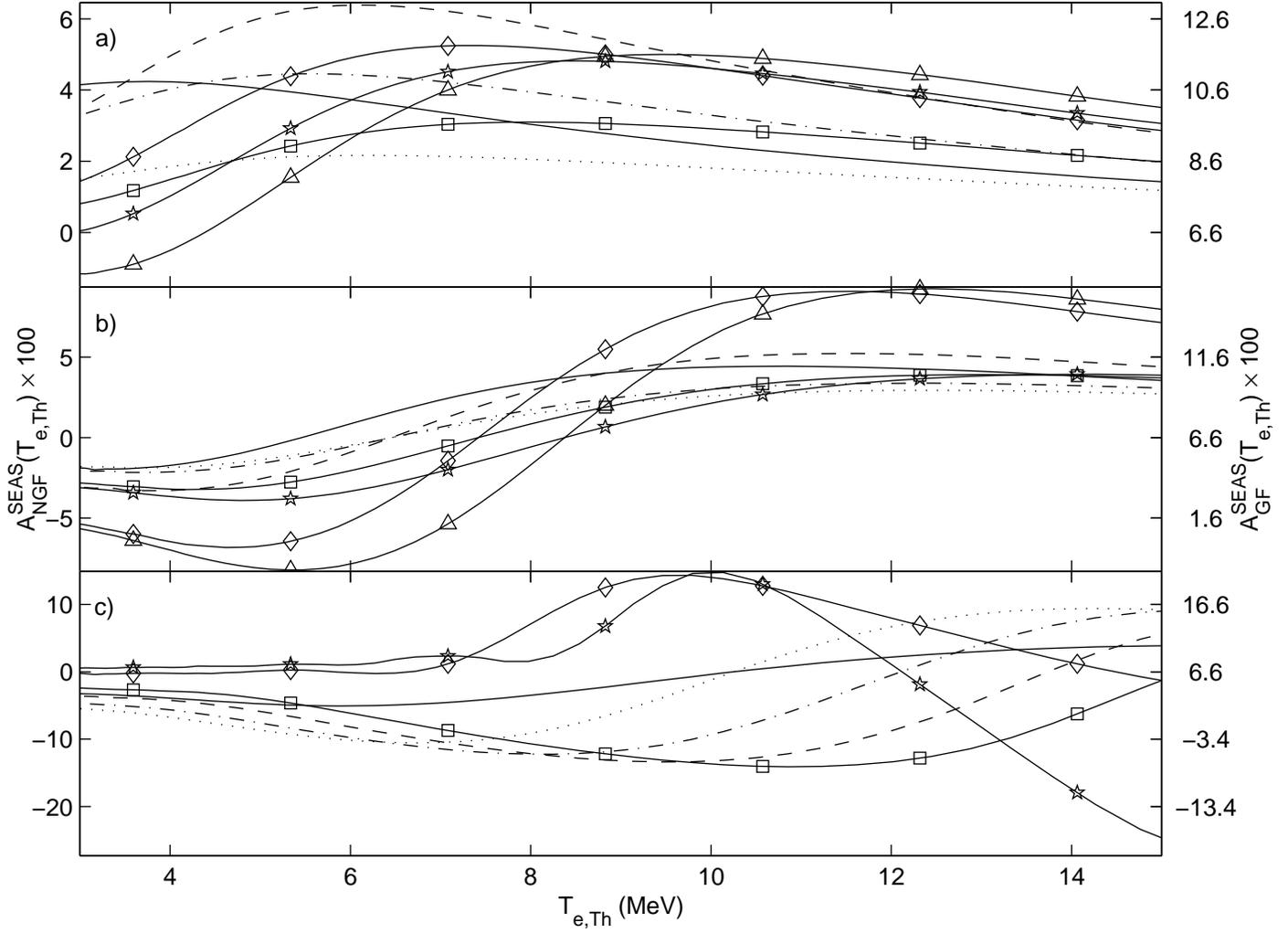
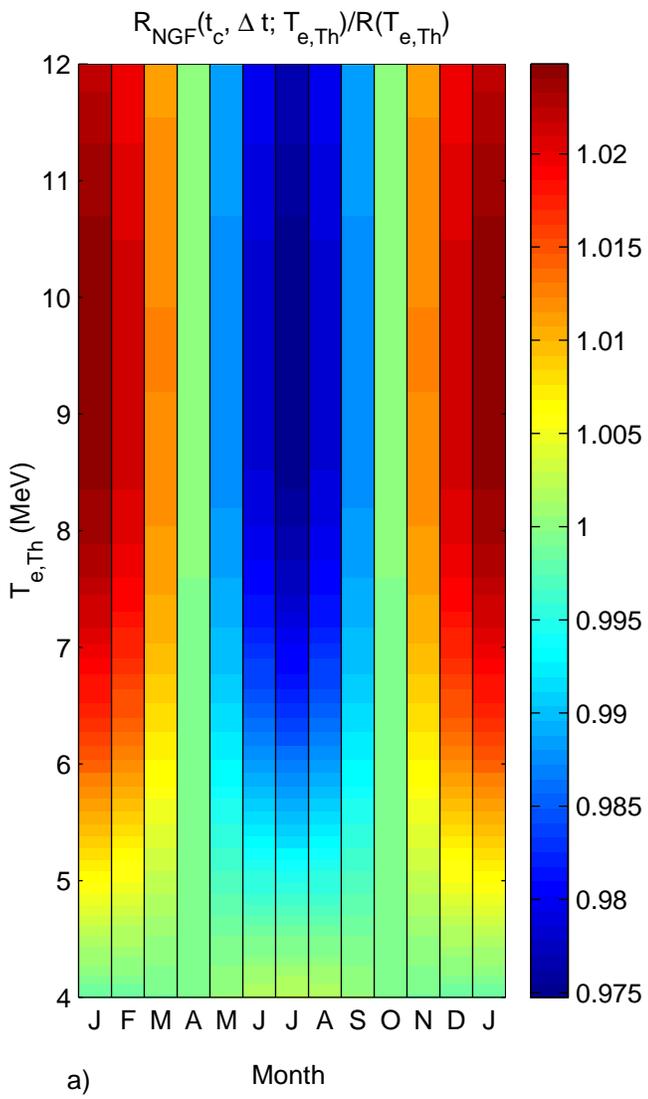
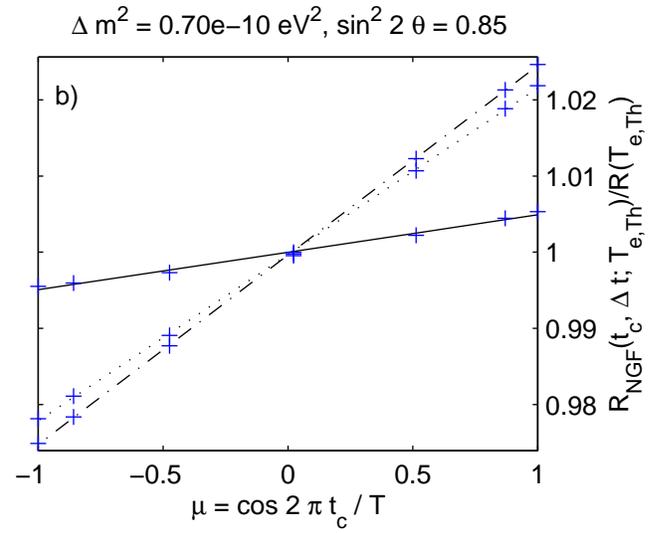
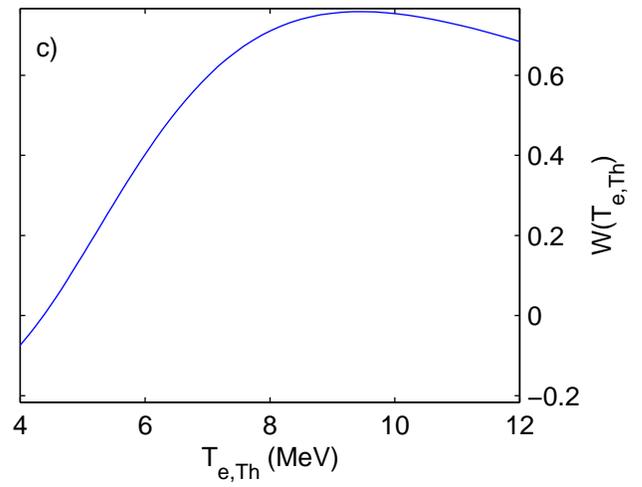
\end{document}